# The Use of ITIL for Process Optimisation in the IT Service Centre of Harz University, exemplified in the Release Management Process

Hans-Jürgen Scheruhn    Christian Reinboth    Thomas Habel

Department of Automatation and Computer Science
Harz University – University of Applied Studies and Research
Friedrichstraße 57-59, 38855 Wernigerode, Germany

{hscheruhn, creinboth, thabel}@hs-harz.de

**Abstract**

This paper details the use of the IT Infrastructure Library Framework (ITIL) for optimising process workflows in the IT Service Centre of Harz University in Wernigerode, Germany, exemplified by the Release Management Process. It is described, how, during the course of a special ITIL project, the As-Is-Status of the various original processes was documented as part of the process life cycle and then transformed in the To-Be-Status, according to the ITIL Best Practice Framework. It is also shown, how the ITIL framework fits into the four-layered-process model, that could be derived from interviews with the universities IT support staff, and how the various modified processes interconnect with each other to form a value chain. The paper highlights the final results of the project and gives an outlook on the future use of ITIL as a business modelling tool in the IT Service Centre of Harz University. It is currently being considered, whether the process model developed during the project could be used as a reference model for other university IT centres.

## 1   Introduction

### 1.1   ITIL

The IT Infrastructure Library (ITIL) is a Best Practice Framework developed by the British Central Computer and Telecommunications Agency (CCTA[1]) for the UK government during the 1980s, and made widely available by the Office of Government Commerce (OGC[2]). It allows the definition, evaluation and optimisation of various business-related IT processes. The ITIL framework is in constant development and has been established as a worldwide de-facto standard for business frameworks in the IT area during the course of the last decade (BSI).

The focal point of all systematic process descriptions contained in the framework, is the economic fulfilment of the underlying business requirements. The goals of ITIL are: increasing the customer satisfaction level as well as the general productivity (especially the more productive and targeted use of existing know-how) and improving communication between IT support and service personnel and customers or colleagues. The entire ITIL knowledge is publicly available in a 40-book library of ITIL publications. ITIL is the only existing universal, public and non-proprietary process framework in the IT sector (itSMF).

Since economical goals and ideas are of general importance in the course of implementing service processes, ITIL is gaining importance as well. A recent study reveals, that in the midst of 2004, 80% of all key decision makers had already concerned themselves with ITIL, 30% of all businesses surveyed had already implemented ITIL structures (COMPASS 04). It is remarkable, however, that not every of the eight ITIL core processes has received the same attention in the business world. Incident Management implementations are especially common, followed by the Change Management, the Problem Management and the Service Level Management as described in this paper (COMPASS 06).

Although the ITIL framework leaves the user a lot of freedom as to the modus operandi of the implementation of ITIL guidelines on the detail level (as outlined in 4), and despite the criticism that this freedom also opens up the possibility of mistakes and misinterpretations, leading experts of the German ITIL movement, organised in the itSMF e.V. ITIL Association, agree: although businesses as well as public institutions have to assess the individual value of ITIL for their own use, the latest improvements to the ITIL framework will ensure it's continuous success (Schmidt and Pränger, 2005).

### 1.2   Project Goals

Harz Universities own ITIL project team, consisting of the authors of this paper and a number of students[3],

---

[1] http://www.ccta.gov.uk

[2] http://www.ogc.gov.uk/index.asp?id=2261

[3] Benjamin Bock, Janine Stiefel, Diana Lamm, Gregor Jagodzinski, Christof Szwarc

documented various processes in the university-owned IT Service Centre as part of an online process management coursework. Two of these processes, the Incident Management and the Release Management, were then remodelled according to the ITIL framework. The project goals were: testing the applicability of ITIL for the use in smaller universities as well as other institutions of higher education and supporting the ongoing quality improvement programme of Harz University by adding to the quality of various IT Service Centre processes.

A recent poll (November 2005) among HS Harz students revealed, that the IT Service Centre is doing a good job already: concerning the reaction time of the centre to requests and problems, almost 30% of all students polled are "very satisfied" with the performance of the centre, an additional 50% describe themselves as "satisfied".

Nonetheless, the universities' administration believes improvement is possible and achievable through a process optimisation according to the guidelines contained within the ITIL framework. A repeat of the student poll mentioned above will take place in 2007 after the implementation of every updated process is completed.

This project can thus be seen as part of an internal assessment on whether the ITIL framework can successfully be used to increase the performance of IT services within a relatively small (3000+ students) public university.

### 1.3 Relation to the Process Life Cycle

The project began with the development of a Balanced Scorecard (BSC) in a conjoint effort made by the ITIL project team and staffers from the IT Service Centre. The general focus of this BSC lay on the long-term development and the continuous improvement of the support quality offered by the centre.

The BSC is a holistic management method, developed in the 1990s by Robert Kaplan and David Norton. It balances the financial perspective – erroneously too often the only perspective that is examined in greater detail without the BSC approach – with the customer perspective, the internal process perspective and the so-called learning & growth perspective that encompasses various aspects of personnel development and motivation (Kaplan and Norton, 1997).

Staffers from the IT Service Centre performed the necessary identification and description of the processes; the As-Is-Status of these processes was then modelled in cooperation with the ITIL project team. The process analysis as well as the process optimisation – ending in the modelling of the To-Be-Status of said processes – was performed by the project team as well. The updated process models are currently being implemented by the IT staff, utilizing software such as ORTS (see 5) Suitable measures for process controlling have yet to be specified – but aside from that, the project described here touches every aspect of the Process Life Cycle (as shown in fig.1).

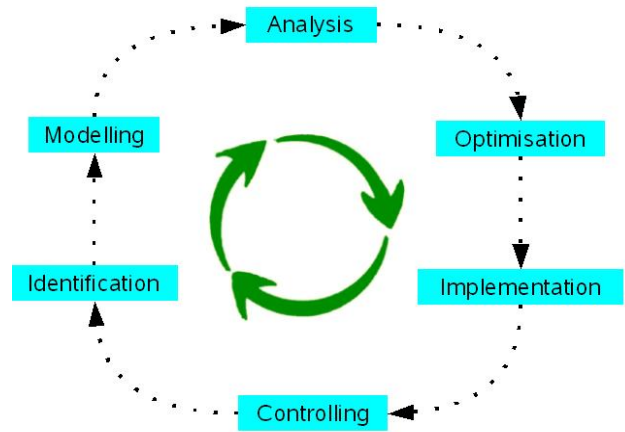

Figure 1: Process Life Cycle

## 2 The Release Management at Harz University

### 2.1 Release Management Tasks

The Release Management is directly responsible for the implementation of changes in existing hard- and software environments. The goal of these measures is to introduce change to productive systems in a manner that causes minimal problems, disturbances and interruptions. Therefore, the first thing that needs to be done is to specify the decision criteria, by which a release change is to be initiated. Furthermore, the consequences of such a change need to be evaluated, the involved hardware and software objects have to be selected and all users have to be notified of the planned changes and prepared for possible problems (SMF).

The main tasks of the Release Management can be summarized as follows:

- Evaluation of new software/file versions
- Building of complete release packages
- Specification of release policies
- Testing of pre-release versions
- Deployment of new releases
- Administration of master copies
- Definition of documentation standards
- Measuring of user acceptance

In ITIL, the Release Management is part of the Service Support. There are interactions with the Change Management (approval of planned changes) as well as with the Configuration Management (changes are entered into the Configuration Management's database) (SMF).

### 2.2 The Release Management Value Chain

During the documentation of the As-Is-Status, the processes that were documented in the IT Service Centre were assembled into a value chain (as described by Porter, 1985) with six phases / basic activities:

1. Concept and Planning
2. Arrangement and Compilation
3. Testing
4. Installation Preparation
5. Distribution and Installation
6. Go Live

The final phase within this chain – the Go Live phase – simply depicts the handover of the updated system to the user and represents the economical conclusion of the Release Management Process as a business process. This phase cannot be modelled, because it contains no actions and only one event, therefore only the sub-processes from the first five phases were modelled using ARIS (see 2.3).

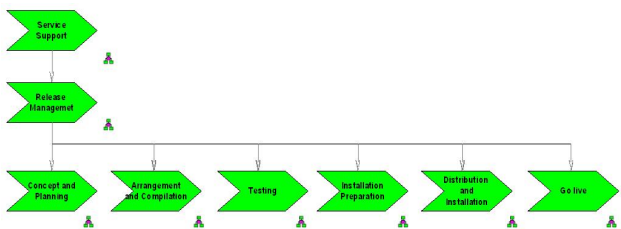

**Figure 2: The As-Is-Status of the Release Management Value Chain**

The value chain described here can be seen in fig.2. It should be noted that this chain contains two additional segments, right above the concept and planning phase – the first of the six phases. These two segments show that the Release Management's value chain is part of the general Release Management Process, that itself is part of the Service Management Support.

## 2.3 As-Is-Status of the Release Management

The modelling of the As-Is-Stati was done utilizing so-called Event-Driven Process Chains (EPCs, see fig. 4). EPCs are graphic representations of business processes and a substantial component of the IDS Scheer AG's[4] ARIS[5] concept. Within EPCs, objects are interconnected with each other in so-called digraphs (the word "digraph" is an abbreviation of the term "directed graph") via lines and arrows in one-to-one allocations, whereby events and functions always have to alternate. Aside from these two objects, an EPC can also contain connectors (such as the binary operators OR, XOR and AND) as well as information objects connected to the functions. These information objects can contain a wide variety of extra information concerning certain functions, e.g. what person (or role) is in charge of a specific function or when and where results are to be documented. (Scheer, 2001).

### 2.3.1 Concept and Planning

During the first process phase, problem reports are received and evaluated. If a decision is made, to further process the report, all available solution models are determined and examined on the basis of their respective hardware requirements and the release environment. At the end of this phase, the staff member in charge of operations decides on a pre-selection of possible solution models for further evaluation.

### 2.3.2 Arrangement and Compilation

In this phase, the necessary instructions and documentations for the installation and the configuration of the new release are gathered and compiled, thus creating the conditions for a successful test of the release, to be conducted in the next process phase.

### 2.3.3 Testing

During the test phase, all solution models that were pre-selected in the first process phase and prepared in the second, are tested parallel to each other. The test results are evaluated and compared afterwards, leading to a final decision for the best possible solution model.

### 2.3.4 Installation Preparation

After the test phase is completed, the preparation of the installation phase can begin. The function tree (see fig. 3) shows that this phase consists of three sub-phases, not all of which have to be completed or even initiated to reach the next phase. Each of these sub-processes is represented by a separate EPC. Depending upon complexity and extent of the release change, there are two possible release modes: the setup of a separate testing system for the purpose of creating an environment for release tests or the simple configuration of software and data in case of less complex releases and driver updates.

The role name for the supervising staffer during the setup process for the testing system has also been altered – from the general ARIS role "computer technician" to "test manager" (see EPC in fig. 4). This corresponds with the conventions for role names under ITIL guidelines. These role names allow the quick identification of the respective persons responsible for every aspect of any process (ITIL).

---

[4] http://www.ids-scheer.de

[5] ARIS = Ger. – Architektur integrierter Informationssysteme – Architecture of integrated information systems. The idea behind this concept, as devised by Prof. A. Scheer, is finding the optimal way of adapting business information systems to their requirements.

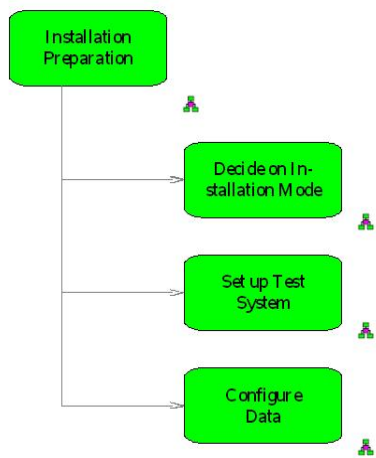

Figure 3: Function Tree of the Installation Preparation Process

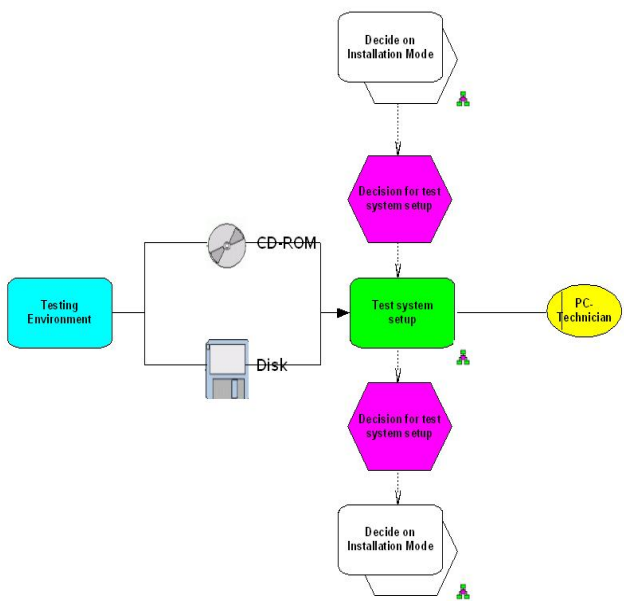

Figure 4: As-Is-Status of the EPC "Setup of a testing system"

### 2.3.5 Distribution and Installation

After the testing system has been prepared, the data is transferred to the selected systems. The decision between separate installation processes for each system and a cloning process is made based upon the total number of selected systems. In case an incorrect transmission of data occurs, the first follow-up step is a problem analysis followed by either a new attempt at an automatic installation or the initiation of a manual installation. If both measures remain unsuccessful, the Problem Management Process is initiated.

## 3 Process Optimisation with ITIL

### 3.1 The use of ITIL

The transformation of the identified release management processes was implemented utilizing the ITIL frameworks for the Service Support process level (ITIL). ARIS was likewise used for modelling the To-Be-Stati (see 2.3).

### 3.2 To-Be-Status of the Value Chain

The fundamental change in the value chain (previously described in 2.2) lies in the now continuous documentation implemented in all processes and sub-processes, a documentation that was inconsistent in the As-Is-Status of the value chain. The complete and continuous documentation of all decisions made and all measures taken corresponds with the ITIL best practice guidelines. The documentation simplifies and therefore facilitates future conversions of Release Management Processes because if gives staffers the option of accessing information on past experiences, problems and the solutions found.

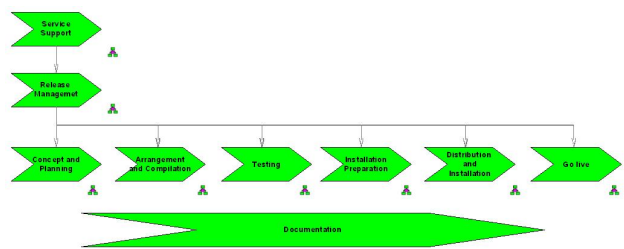

Figure 5: The To-Be-Status of the Release Management Value Chain

### 3.3 To-Be-Status of the Release Management

#### 3.3.1 Concept and Planning

A thorough comparison of the As-Is-Status of the concept and planning phase with the ITIL guidelines showed little to no potential for optimisation. Only the now continuous documentation, now part of every process in the entire value chain (see 3.2), has been integrated into the EPC.

#### 3.3.2 Arrangement and Compilation

Again, the comparison of the As-Is-Status and the ITIL guidelines showed little to no potential for optimisation. Therefore, only changes concerning the level of documentation were made. The relatively low number of changes is, in itself, not an unusual result. On the contrary, it only demonstrates the "best practice character" of ITIL: not every process analysis uncovers great potential for optimisation, especially in processes that underwent an often year-long, internal "natural optimisation" and thus come relatively close to the ideal ITIL processes without any remodelling.

### 3.3.3 Testing

The addition of further documentation procedures has more significant effects in this process phase, since a larger number of steps and decisions need to be documented. Additionally, at the end of the EPC, there is an explicit split into a decision between the setup of a testing system or the simpler configuration of software and data.

### 3.3.4 Installation Preparation

Once remodelled according to the ITIL guidelines, this process phase exhibits multiple and extensive changes, especially concerning the setup of a testing system. Aside from the additional documentation (as described in 3.2), one can find a newly implemented method to reconfigure multiple computer systems using a so-called "virtual user". A virtual user can supervise the setup of new systems during downtime – especially over night, so that the newly reconfigured systems are already available to the user the following morning.

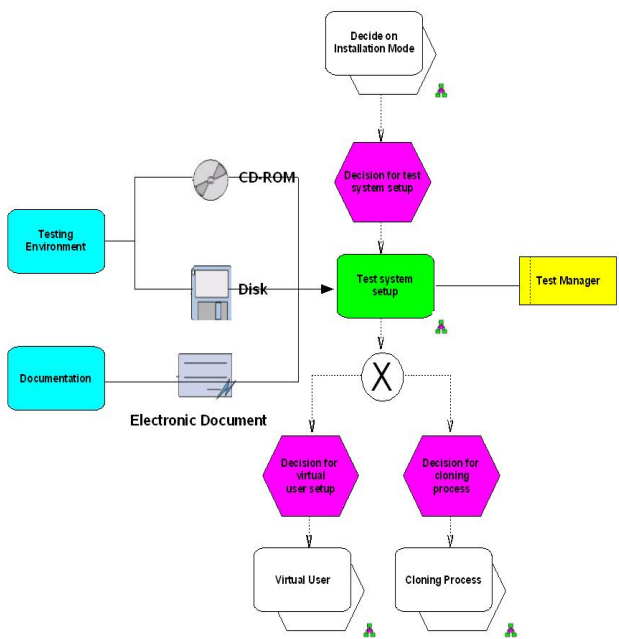

Figure 6: To-Be-Status of the EPC
"Setup of a testing system"

### 3.3.5 Distribution and Installation

The To-Be-Status of the last process phase that can be modelled (the Go Live phase can not, because it represents only the final handover to the user, see also 2.2) was augmented with two new functions: the creation and the configuration of virtual users for testing purposes and the option to completely restore the entire system in the event of a major crash event or a fatal failure during the sometimes-critical release process. The EPCs for the installation and the cloning sub-processes remain unchanged, aside from the additional documentation. One other novelty is, that the process for the manual installation can now be initiated by the virtual user via the option "restore previous configuration". Also, the newly configured computer is not directly handed over to the user after the cloning process ends, but is instead handed to the virtual user for some final tests.

By utilizing this virtual user, a detailed examination of all major functions of the new release can be initiated directly after the cloning process ends. Should the new configuration be found unworkable, the test process can lead directly into the complete restoring of the previous system configuration. During the final testing phase, the virtual user simulates a "real" user by sequentially running through a pre-configured series of software functions and documenting the results. In the event of a failure, the previous system configuration is restored and the individual installation process initiated. The restoring process itself can also be managed by the virtual user, regardless of the reason for the failed test, so that no IT service personnel needs to be activated.

## 4 The four-layered Process Reference Model

During the course of the interviews, conducted by the members of the ITIL project group with the staff of the IT Service Centre, it became more and more clear, that a three-layered process reference model as described in the ITIL reference libraries (ITIL) would be insufficient, if a detailed process analysis was to follow. Instead, a four-layered model was used, comprised of the three layers of the original ITIL process reference model and one additional level referred to as the "Detail Level" in fig. 7.

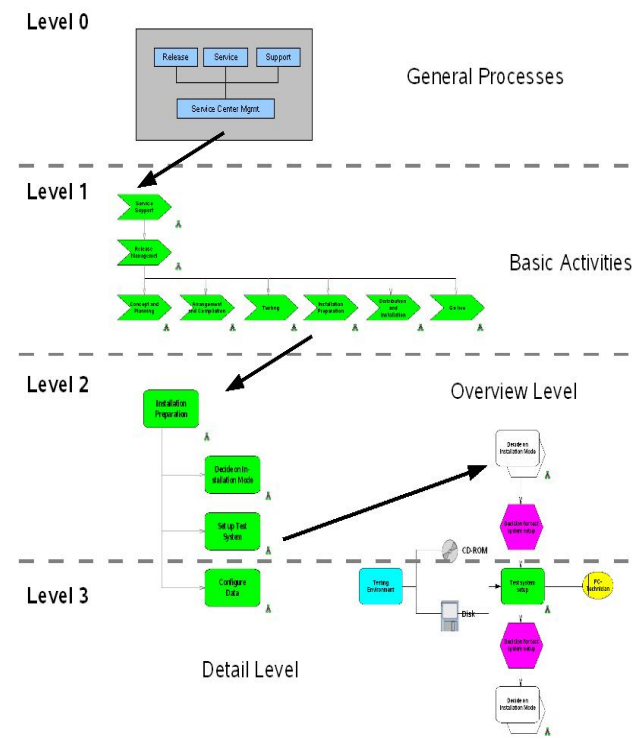

Figure 7: The ITIL Framework covers only the upper three layers of the four-layered model

The upper layer of this model – Level 0 – contains general processes such as the Incident Management or the

Release Management. No graphical representation of any general process has been used in this paper.

In the second layer – Level 1 – we see the division of these processes into basic activities, as shown in the value chain. The value chain for the Release Management Process – as seen in fig. 1 and 4 – is an example for such a "lineup" of basic activities.

The third layer – Level 2, also named the "overview layer" – contains the sub-processes within the respective basic activities from layer one, represented in a function tree (see also fig. 2). The sequence of these sub-processes is not determined by the structure of the function tree and not all processes have to be initiated and/or finished to get to a different function tree.

The detailed steps are shown in the lowest layer – Level 3 – in the form of EPCs. This layer is also designated the "Detail Level" (see fig. 7), because it shows the process workflow in the greatest possible detail without getting too complex to provide oversight. Processes on this level can no longer be modelled on the basis of the ITIL framework but only on the basis of employee interviews. The changes to the processes described in this paper, that were realized in the detail level (namely concrete changes to EPC details) were thus not directly derived from the ITIL framework but rather the realizations of changes that occurred within the superordinate overview level – changes on this level can be directly derived from the ITIL framework.

To give an example on how this realization process works: The ITIL framework's demand for a detailed and more thorough documentation has already been detailed in 3.2. It has been shown how the additional documentation becomes part of the value chain and thus, needs to be implemented in every sub-process that is part of any of the basic activities within the value chain. So, the additional documentation seen in the EPC in fig. 6 is simply the realization of this implementation in the detail level. The modus operandi of this implementation – that is, how this additional documentation should be integrated into the process – can not be derived from the ITIL framework, although the ITIL publications (as mentioned in 1.1) offer helpful suggestions.

## 5   Conclusion and Outlook

It is generally believed, that ITIL will become even more popular and widely used in the coming years, in the business as well as in the public sector. The value of ITIL for technical planning, contribution and support of IT services in the public sector became apparent in the promising outcome of the ITIL project at Harz University. Other work in this field underlines this theory, e.g. the "ITIL simulation game" especially designed for IT service personnel in universities, hosted by Duisburg-Essen-University[6].

The IT Service Centre has accepted the remodelling proposals made by the ITIL project group and is already working on the implementation. Some early tests with the Open Ticket Request System Software[7] (ORTS), that allows the ITIL conform implementation of the Incident Management that was also modelled by the project group, were very successful. A continuation of the ITIL project over the coming semesters is already in planning. The future use of a generic model reference system is also in discussion.

One substantial discovery during the course of the implementation process should also be noted: ITIL can only be used to describe and optimise the upper three process layers in the four-layered model (see also fig. 7) that can be derived from the ITIL framework (ITIL).

In conclusion, it has to be determined, that it is clear that additional work will be required, before the success of the process optimisation efforts can be evaluated in actual metric terms.

Firstly, a process controlling has to be established to complete the Process Life Cycle and to allow the assessment of the updated processes, as outlined in 1.3.

Secondly, the students (as the primary "customers" of the IT Service Centre) will be questioned in early 2007 as part of an ongoing quality management measure about their satisfaction with the IT Service Centre's general performance. A previous poll, as outlined in 1.2, revealed an already high level of satisfaction. It is believed that the optimisation of various performance-related processes through ITIL, as described in this paper, will lead to an increase in customer satisfaction, although it remains to be seen just how much of an impact the changes will have.

Once the polling results are in, this can be determined utilizing the standard methods of statistical analysis, e.g. variance analysis. This will allow the final evaluation of the success of the ITIL project. It is hoped, that the expected increase in satisfaction will stimulate further exploration of the possibilities, which ITIL offers for the optimisation of other, yet unanalysed IT Service Centre processes.

## 6   Acknowledgements

We want to thank Benjamin Bock of the ITIL project group for his efforts and Friedemann Hass, the director of the IT Service Centre, for his continuous support and encouragement. We also want to thank the staffers who sacrificed scarce downtime at work in order to be interviewed. Additional credits go to Andrea Roth and Stefan Schneider, who organised and supervised the student polling in 2005.

---

[6] http://www.wip.uni-duisburg-essen.de/team/bauerschmid

[7] http://www.orts.de/de/